\documentclass[superscriptaddress,aps,twocolumn,prd,showpacs,nofootinbib]{revtex4-1}
\usepackage{amsmath}
\usepackage{amsfonts}
\usepackage{amssymb}
\usepackage{graphicx}
\makeatletter
\newcommand*{\rom}[1]{\expandafter\@slowromancap\romannumeral #1@}
\makeatother

\usepackage[usenames,dvipsnames]{xcolor}
\usepackage{hyperref}
\hypersetup{colorlinks=true,
citecolor=MidnightBlue, linkcolor=MidnightBlue, urlcolor=MidnightBlue}
\usepackage{tikz}

\usepackage{comment}

\newcommand{\be}{\begin{equation}}
\newcommand{\ee}{\end{equation}}
\newcommand{\bea}{\begin{eqnarray}}
\newcommand{\eea}{\end{eqnarray}}

\newcommand{\bseq}{\begin{subequations}}
\newcommand{\eseq}{\end{subequations}}

\allowdisplaybreaks[1]
\usepackage{amsmath}

\newcommand{\orcidicon}{%
	\begin{tikzpicture}
	\draw[lime, fill=lime] (0,0)
		circle [radius=0.16]
		node[white] {{\fontfamily{qag}\selectfont \tiny ID}};
	\draw[white, fill=white] (-0.0625,0.095)
		circle [radius=0.007];
	\end{tikzpicture}	\hspace{-2mm}
}

\newcommand\orcidIsmael{{\href{https://orcid.org/0000-0002-0606-764X}{\orcidicon}}}
\newcommand\orcidRuth{{\href{https://orcid.org/0000-0001-5536-3130}{\orcidicon}}}

\begin{document}


\title{Characterization of 
wormhole space-times supported by a covariant
action-dependent Lagrangian theory}

\author{Ismael Ayuso\orcidIsmael\!\!}
\email{ismael.ayuso@ehu.eus}
\author{Ruth Lazkoz\orcidRuth\!\!}
\email{ruth.lazkoz@ehu.es}
\affiliation{Department of Theoretical Physics, University of the Basque Country UPV/EHU, P.O. Box 644, 48080 Bilbao, Spain}

\date{\today}

\begin{abstract}

In this work, we undertake an analysis of new wormhole solutions within an action-dependent Lagrangian framework. These geometries can be traversable and supported by a positive energy density. The modification of the gravitational field equations is produced by the inclusion in the gravitational Lagrangian linear  of a 
background four-vector $\lambda_{\mu}$. This new term expands significantly  the conventional description of gravity making it highly non-linear, and therefore drawing general conclusions about legitimate forms of $\lambda_{\mu}$  proves a formidable task in general. It is, then, customary to adopt an ansatz that strikes a balance between enabling new  phenomenology while retaining  a significant degree of generality on $\lambda_\mu$. Ours is given by the choice $\lambda_\mu=(0,\lambda_1(r), 0, 0)$, with an arbitrary $\lambda_1(r)$.  
By setting $\lambda_1(r)=-1/r$ we craft new  families with physically desirable properties, but the wormholes thus generated turn out to be conical, as evidenced by an angle deficit, in a similar fashion to other known solution families. Under the general shape of $\lambda_1(r)$, we demonstrate that these solutions are not compatible with  the Null Energy Condition (NEC) in general, as it happens to their General Relativity counterparts, except on specific occasions where the derivative of the redshift function of the metric diverges at the throat (however, in these latter cases, the traversability of the wormhole will be disrupted). On the other hand, it is possible to solve the conical character and satisfies the flatness condition for more general functions of $\lambda_1(r)$.

\end{abstract}

\maketitle

\section{Introduction}

Wormholes are enticing theoretical constructs with the potential to link disparate regions of spacetime.  In this sense, they can be interpreted as a geometric passageways between connecting events, within the same or across different universes. However, one major departure from other solutions to Einstein's field equations lies in the necessity  to make ad hoc topological considerations to complete their description, as the primary mathematical purpose of the left hand side of Einstein's equations  is just to address geometry.

The foundation stone in the field of wormholes was placed by Flamm \cite{Flamm, Flamm_rep} through an original analytic and pictorial exploration of the then recent Schwarzschild solution in 1916. This theoretical issue remained in standby until 1935, when Einstein and Rosen revived it  by proposing an atomistic theory of matter and electricity where a particle would  be represented by a ``bridge'' connecting two identical spacetime sheets \cite{Einstein:1935tc}. 
Subsequently,  the relativistic community   coined the term ``Einstein-Rosen bridge" (ERB) to describe a configuration connecting two points of space to one another by means of a throat.

Again, this field of study lay dormant for two decades, until in 1955 John Wheeler rescued it by studying some topological aspects of General Relativity (GR) \cite{Wheeler:1955zz,Wheeler:1957mu}. He hypothesized about a particle with a ``gravitational electromagnetic entity" which he called  ``geon".  In modern language it could be seen as an ``unstable gravitational-electromagnetic quasisoliton".  Thus, geons emerged as solutions in a multiconnected spacetime, where two widely separated regions were connected by a tunnel or bridge in the context of the coupled Einstein-Maxwell field equations. A follow-up study of this model, together with an analysis of Riemannian geometry of manifolds of nontrivial topology, was presented by Wheeler and Misner in \cite{Misner:1957mt}. This work represented an attempt to unify and elucidate all related physics, and it mark the inaugural use of   the term ``wormhole". A sentence that encapsulates their view is this: "Physics is geometry", since all physical phenomena emerge as manifestations of the underlying geometry.

After that,  attempts to build tranversable wormholes, including the studies of Ellis \cite{Ellis:1973yv} and Bronnikov \cite{Bronnikov:1973fh}, reached a pinnacle in the astounding and instructive paper of Morris and Thorne \cite{Morris:1988cz} (see as well \cite{Lemos:2003jb} for a review). Unfortunately, the matter that bolsters these  solutions is typically exotic, in the sense that it violates most energy conditions.  However, some wormholes solutions are known which try to circumvent the requirement of extravagances in the matter content, or at least minimize its presence. A comprehensive reference where this is  discussed is \cite{Visser:1989kh}. Much more activity ensued based on those interesting problems and it lead to  Visser's lesson compilation  \cite{Visser:1995cc}, which  distills a vast amount of knowledge about wormholes. For all these reasons, wormholes represent fabulous laboratories offering unparalleled routes to the understanding of physics of gravitation in theories beyond GR \cite{Lobo:2009ip, Harko:2013yb, Boehmer:2012uyw, Capozziello:2012hr, BeltranJimenez:2017doy, Elizalde:2023rds, Gonzalez:2008xk, Gonzalez:2008wd, Houndjo:2013us, Rastgoo:2024udl, Banerjee:2021mqk, Bubuianu:2024zsm, Mustafa:2024qts}.

One of such new proposals acts as catalyst of our investigations, as we are going to be concerned  specifically with wormholes in non-conservative theories of gravity. The principle of matter-energy conservation is one of the pillars of GR, however,  its applicability becomes a subject of debate within covariant gravitational theories. This issue  traces back to the debate between  Klein, Noether,  Hilbert and Einstein concerning the mathematical relevance of the conservation of matter-energy \cite{Brading:2005ina}, and was analyzed throughout the 20th century in the literature \cite{PhysRev.111.315, PhysRev.112.287, PhysRev.113.934,Bondi:1990zza, Bak:1993us}. 

Moreover, this conservation law implies two important features in GR:  diffeomorphism invariance on the one hand and minimal coupling between gravity and matter on the other. For that reason, we can explore nonconservative theories from a bunch of different perspectives. One such possibility consists in breaking conservation by hand as in Rastall gravity \cite{Rastall:1972swe}. We may also eliminate either diffeomorphism invariance or minimal coupling to explore nonconservative gravity. For example, the Brans-Dicke scalar-tensor theory \cite{Brans:1961sx}, in which a scalar field appears non-minimally coupled to gravity in the Jordan frame, shows that when working in the Einstein frame, the dilaton (coming from the conformal transformation) becomes minimally coupled to the curvature instead of appearing coupled to the material action, thus causing the conservation law to hold in very specific cases. The covariant conservation of energy-momentum in modified gravities with more intricate non-minimal couplings is presented in the paper \cite{Koivisto:2005yk}.

Another possibility is to apply the Herglotz problem to the field of gravity to make the Lagrangian density depend on the action itself, thus breaking diffeomorphisms invariance. A realization of this approach can be found in the recent theory of Lazo et al, (both in its non-covariant form \cite{Lazo:2017udy} and in its covariant form \cite{Paiva:2021kuk}). In this proposal non-conservation is caused by the presence of a background four-vector that introduces a privileged direction into the theory. 
These intriguing settings and other examples such as those above, along some of their fundamental aspects in the non-conservation framework, have been  pedagogically examined in the work \cite{Velten:2021xxw}.

Summarizing, a modified gravity setting arising from a non-conservative gravitational theory has the ability to introduce new interesting phenomenology \cite{Fabris:2017msx}. For the present work, we will focus on the latter mentioned case proposed by Lazo et al. as a generalization of the Herglotz problem within the gravitational framework  \cite{Lazo:2017udy, Paiva:2021kuk}. An important difference between those two works is the absence of a surface term in the Lagrangian, which is not allowed in the Herglotz variational problem \cite{Paiva:2021kuk}. This lead to  an initial proposal with non-covariant field equations which clearly was not completely acceptable. A more careful analysis including the required surface term allowed to  obtain the covariant field equations in \cite{Paiva:2021kuk} for which we produce  wormhole solutions. The conceptual image of this non-conservative setting is that of a theory in which the geometry resembles a non-perfect elastic rubber sheet, i.e. there is a dissipative effect, but it is originated by the geometry itself. 

The program we initiate in this paper is the characterization of transversable wormholes that can be found in this modified gravity theory. The steps of our construction are outlined in detail in the hope that this will pave the way to additional solutions in the future. We also delve into the general features of the spacetimes thus built; in particular, we explore questions such as asymptotic flatness, the Null Energy Condition (NEC), energy conservation (or violation) or pressure gradiants. This gives us an idea of the array of unwanted consequences that may arise when considering a broader theoretical framework in order to obtain a solution with some basic desirable features. 

Let us now offer a guide to the organization of this manuscript. In Section \ref{II} we will produce a convenient introduction to the state of the art and mathematical tools of wormholes. Following this, in Section \ref{III} we present in detail the framework of the generalized non-conservative theory of gravity that we use. After those two preliminary sections, the equations of motion of the theory for a static and spherically symmetric wormhole are presented in Section \ref{IV}. In fact this can be viewed as a revision of the two earlier sections when we introduce a specific energy-momentum tensor. At this point, the model is described up some specific functions of the metric or the profiles of the energy density and pressure. Results in this section are useful to illustrate  dissipative character of the theory later in section \ref{V}.  In Section \ref{VI}, we start from the easy case of Schwarzschild wormhole in order to obtain more complex (and traversable) solutions in Section \ref{VII}, upon fixing the remaining degrees of freedom. This will lead us to study in Sections \ref{VIII} and \ref{IX} the main issues that arise, namely, the violation of the NEC and the conicity character of wormholes, respectively. Finally we summarize our work in Section \ref{X}.

\section{State of the art of wormholes in General Relativity}\label{II}

The purpose of this section is to review some relevant aspects that lay the groundwork for the subsequent analysis. In this sense, we start by presenting one of the most famous wormhole metrics in literature, the Morris-Thorne metric \cite{Morris:1988cz}:

\begin{equation}
    ds^2=-\displaystyle e^{2\Phi(r)}dt^2+\frac{dr^2}{1-\frac{b(r)}{r}}+r^2(d\theta^2+\sin^2\theta d\phi^2)\; . 
    \label{MTst}
\end{equation} 

This represents a static and spherically symmetric spacetime, and, in addition, we assume it represents a solution to the Einstein field equations which will eventually be completely determined by stating $b(r)$ and $\Phi(r)$.

It is imperative now to discuss what is required for 
Eq. \eqref{MTst} to represent the line-element of a wormhole, and in particular a trasversable one. Note that such characteristic is not guaranteed a priori; actually, that metric can accommodate the Minkowski space-time, or the Schwarzschild wormhole, which is neither  trasversable nor stable.

In this context $\Phi(r)$ is termed the redshift function as it is associated with the gravitational redshift, thus  playing a pivotal role towards trasversability. In fact, the absence of horizons is a must for the wormhole to be trasversable, which implies that $g_{tt}=e^{2\Phi(r)}\neq 0$, or in other words, $\Phi(r)$ be finite  \cite{10.1063/1.1664717, Carter1969KillingHA, Chrusciel:1998rw}. The influence of $\Phi(r)$ on the reddening of photons is linked to the derivative of that function with respect to $r$, as it determines the strength of the attractive or repulsive nature of gravity \cite{bookLobo}, and also the intensity of the tidal forces. This aspect becomes evident when the tidal tensor is defined as $E^{i}_{\;\;j}=R^{i}_{\;\;tjt}$ \cite{Voicu:2011by},  where its non-null components depend on radial derivatives of $\Phi(r)$ , and vanishes if a constant $\Phi(r)$  is imposed\footnote{The non null components of the tidal tensor are:
\bea
R_{trtr}&=&\frac{1}{2} e^{2 \Phi (r)} \left(\Phi '(r) \left(\frac{b(r)-r b'(r)}{r^2-r b(r)}+2 \Phi'(r)\right)+2 \Phi''(r)\right)\nonumber\\
R_{t\theta t\theta}&=&(r-b(r)) e^{2 \Phi (r)} \Phi '(r) \nonumber\\
R_{t\phi t\phi}&=&(r-b(r)) e^{2 \Phi (r)}\sin^2{\theta} \Phi '(r)
\eea}.

In turn $b(r)$ is referred to as the shape function, as it determines the unique appearance of the wormhole. As a matter of fact, visualizing the shape of the wormhole is one of the customary facets of study of specific wormhole spacetimes.  Embedding diagrams are helpful in this task, thus warranting more investigation.

First, staticity and spherical symmetry allow us to choose a  fixed time and $\theta=\pi/2$  slice,  without loss of generality. Our purpose is to   embed that two-dimensional surface in the three-dimensional Euclidean space. This approach lets us express the intrinsic curvature  of the slice as an extrinsic curvature of a two-dimensional surface within a space with an extra dimension.

Upon specifying that particular slice the line element simplifies to 
\bea
ds^2=\frac{dr^2}{1-\frac{b(r)}{r}}+r^2 d\phi^2,
\label{emb1}
\eea
and it can be compared with that of three-dimensional Euclidean space written in cylindrical coordinates:
\bea
ds^2=dz^2+dr^2+r^2d\phi^2 \; .
\label{emb2}
\eea

By harmonizing those two line elements we get the embedding equation
\bea
\frac{dz}{dr}=\pm\left(\frac{r}{b(r)}-1\right)^{-1/2}\; ,\label{embeddingequation0}
\eea
which leads to
\bea
ds^2=\left(1+\left(\frac{dz}{dr}\right)^2\right)dr^2+r^2d\phi^2\; .
\label{embeddingequation}
\eea
The double sign in Eq. \eqref{embeddingequation0}  gives us, in fact, two hypersurfaces which are  mirror images of one another. Their boundaries are given by the condition 
$b(r_0)=r_0$ which defines a minimum radius $r_0$ obviously associated to $dz/dr \rightarrow \infty$. Those two surfaces are then glued together at their boundaries and we get the defining wormhole throat, which is a geodesically complete multiply connected hypersurface.

Besides, we should demand that the constructed spacetime be asymptotically flat, and  this implies both $b(r)/r\to 0$ (that is, $dz/dr\to 0$) and $\Phi(r)\to 0$ as $r\to \infty$. In addition, continuing with the embedding scheme, a wormhole must adhere to the flaring-out condition to guarantee that the throat outspreads. This condition comes from  analyzing the form of the function $r(z)$, which should have a minimum at $z=0$ (the throat), i.e:
\bea
\left.\frac{d^2r}{dz^2}\right|_{\text{throat}}>0. \label{floutthro}
\eea
At the end of the day, this just means the curve $r(z)$ is concave at the throat.  This can be understood by considering that the shape of the embedding  surface should resemble a parabola, and thus, sufficiently far from $z=0$, ${d^2r}/{dz^2}$ will be a positive constant, and the condition of Eq. \eqref{floutthro} can be extended to all $r$. The limiting case is that of the hyperbola, where far enough from the throat, the second derivative becomes zero as $z$ and $r$ grow at the same rate (this would be the case of a conical wormhole). 
Mathematically for the metric used, this result follows from the differentiation of the inverse equation of Eq. \eqref{embeddingequation0}:
\bea
\frac{d^2r}{dz^2}=\frac{b(r)-b'(r)r}{2b^2(r)}>0\ .\label{floutcondition}
\eea
Moreover, the latter can be translated into the simple requirement $b'(r_0)<1$ at the throat, where the flaring-out condition is translated again into the condition to have a minimum at the throat in $r(z)$. This will have important consequences for the fulfilment of energy conditions.

Let us to remember, at this point in our discussion, that no theory of gravitation has been proposed so far  and that we are just talking about the geometry of spacetime used. For this reason the discussion of this section stands totally valid for a spacetime of the form of Eq. \eqref{MTst} irrespective of considerations involving the gravitational Lagrangian.

\section{Theoretical framework of a generalized nonconservative gravity}\label{III}

The second pillar of this work comes from the generalized variational problem of Herglotz. The starting point of this is the Action Principle, introduced by Euler, Hamilton, and Lagrange, which is the foundation stone of all physical theories. According to it, theoretical frameworks begin with the specification of  a Lagrangian and the action derived from it. This, in turn, serves as the basis for deriving the dynamical equations.

Building upon these grounds, Herglotz pondered the possibility of letting the Lagrangian not only of the usual variables but also of the action itself. The conclusions to this riddle in classical mechanics are presented in \cite{Herglotz1, Herglotz2}, along with contemporary efforts to generalize the Herglotz principle for multiple independent variables \cite{2003JMP....44.3911G}.  The starting point is to consider a Lagrangian structured as follows:
\bea
\mathcal{L}\equiv \mathcal{L}(x,\dot{x},S,t)\; ,
\eea
where $S$ must satisfy:
\bea
\dot{S}=\mathcal{L}(x,\dot{x},S,t)\; , \label{clasmec}
\eea
with the dot denoting a time derivative. Subsequently, using the usual variational principles of classical mechanics, Herglotz determined that the generalized Euler-Lagrange equation for the aforementioned actions is \cite{Herglotz2}:
\bea
\frac{d}{dt}\left(\frac{\partial\mathcal{L}}{\partial\dot{x}}\right)=\frac{\partial\mathcal{L}}{\partial x}+\frac{\partial\mathcal{L}}{\partial S}\frac{\partial\mathcal{L}}{\partial\dot{x}}\; .
\eea
This describes a dissipative system because of the presence of the rightmost term. Some interesting examples in classical mechanics are elucidated in \cite{Herglotz2}.

However, in this work we are interested in gravitational settings. Such a generalization of the original discussion within the domain of gravity can be found in \cite{Lazo:2017udy, Paiva:2021kuk}. Let us follow those references to  consider a spacetime defined by an n-dimensional manifold $\Omega$ endowed with a metric $g_{\mu\nu}$, with $\delta\Omega$ being the boundary of $\Omega$ with $h_{\mu\nu}$ as the induced metric in the $(n-1)$-dimensional boundary:
\bea
&\nabla_\mu s^\mu =\mathcal{L}(x^\nu,g_{\alpha\beta},g_{\alpha\beta},_{\nu},s^\nu)\; ,\label{genS}\\
&S(\delta \Omega)=\int_{\delta\Omega}{\sqrt{|h|} n_\mu s^\mu d^{n-1}x}=\int_\Omega{\sqrt{-g} \nabla_\mu s^\mu d^nx}\; .\nonumber
\eea
In the latter the comma denotes partial derivatives, $\nabla_\mu$ is the covariant derivative, and Stokes' theorem is used ($\delta\Omega$ must represent an orientable Jordan surface with its normal vector being $n^\mu$). It becomes then feasible to find the equations of motion from a variational point of view in  a similar way to the Herglotz problem in classical mechanics. 

Following \cite{Paiva:2021kuk}, we  adopt a specific  theory of gravitation based on the following Lagrangian:
\begin{equation}
    \mathcal{L}=\mathcal{L}_g+\lambda_\mu s^\mu+\mathcal{L}_m=R+\lambda_\mu s^\mu+\mathcal{L}_m\; ,
\end{equation}
where $\lambda_\mu$ is an arbitrary vector field playing the role of a cosmological four-vector, and $s^\mu$ is the differentiable action-density vector field. The rationale for introducing the term $\lambda_\mu s^\mu$ into the Lagrangian originates  from the intention to implement the linear dependence of the action, addressed by Herglotz in the context of classical mechanics, in a covariant manner. In this way, this term should be associated with the inherent dissipative character of the theory.

The connection between $S$ and $s^\mu$ is established to provide  a covariant version for the classical equation \eqref{clasmec} that implies the generalization of time derivative of $S$ of \eqref{clasmec} through the change of this time derivative of $S$ to a divergence of a certain auxiliary four-vector $s^\mu$, as it can be seen in \eqref{genS}. Henceforth, the classical action is related somehow with the component $s^0$ as it is discussed in \cite{Lazo:2017udy}. Accordingly, $\lambda_\mu$ assumes the role of a four-vector coupling parameter related to the specific dependence of the Lagrangian on the action.

Assuming the conventional coupling with the matter section, the equations of motion can be calculated from variational principles, thereby extending the Herglotz problem \cite{Paiva:2021kuk}:
\begin{equation}
    R_{\mu\nu}-\frac{1}{2}g_{\mu\nu}R+K_{\mu\nu}=\frac{8 \pi G}{c^4} T_{\mu\nu}.
    \label{eqmotion}
\end{equation}
In the latter we have
\begin{equation}
    K_{\mu\nu}=\Lambda_{\mu\nu}-g_{\mu\nu}\Lambda,
\end{equation}
with $\Lambda_{\mu\nu}$ being the following symmetric tensor:
\begin{equation}
    \Lambda_{\mu\nu}=\frac{1}{2}\left(\lambda_{\mu;\nu}+\lambda_{\nu;\mu}\right)-\lambda_\mu\lambda_\nu,
\end{equation}
of course, $\Lambda=\Lambda^\mu_{\;\;\mu}$. In addition, from now on, we take $G=c=1$. It is noteworthy that  the action density vector field $s^\mu$ does not appear in Eq. \eqref{eqmotion}, leaving $\lambda_\mu$ as the only residual parameter of the model. Indeed, it is this parameter that is accountable for the dissipative properties which this gravitational theory manifests.

\section{Field equations for the Morris-Thorne wormhole}\label{IV}

Alongside the metric, we will include the energy-momentum tensor for an anisotropic fluid compatible with spherical symmetry \cite{letelier,Cho:2017nhx, bookLobo}:
\begin{equation}
    T_{\mu\nu}=\left(\rho+p_t\right)U_\mu U_\nu+p_t g_{\mu\nu}+\left(p_r-p_t\right)\chi_\mu\chi_\nu\ ,
\end{equation}
where $U^\mu$ is the four-velocity, $\chi^\mu$ is the unit spacelike vector in the radial direction, i.e., $\chi^\mu = \delta^\mu_r/\sqrt{g_{rr}}$; $\rho(r)$ is the energy density, $p_r(r)$ is the radial pressure measured in the direction of $\chi^\mu$ , and $p_t(r)$ is the transverse pressure measured in the orthogonal direction to $\chi^\mu$ \cite{bookLobo}. In addition, the most general four-vector, in order to construct $\Lambda_{\mu\nu}$, reads
\begin{equation}
    \lambda_\mu=(\lambda_0(x^i),\lambda_1(x^i),\lambda_2(x^i),\lambda_3(x^i))\; ,\nonumber
\end{equation}
where $x^i$ represents the dependence on the four coordinates, formally, i.e. ${x^i}=(t,r,\theta,\phi)$.

At this stage, one can work out which non-diagonal terms of $K_{\mu\nu}$ must vanish as a consequence of the diagonal character of $g_{\mu\nu}$ and $T_{\mu\nu}$. The non-diagonal terms can be succinctly expressed as:

\bea
K_{01}&=&-\lambda_0 \lambda_1+\frac{1}{2} \left(\frac{\partial \lambda_0}{\partial r}+\frac{\partial \lambda_1}{\partial t}\right)-\lambda_0\Phi'(r)\; ,\nonumber \\
K_{02}&=& -\lambda_0 \lambda_2+\frac{1}{2} \left(\frac{\partial\lambda_0}{\partial\theta}+\frac{\partial{\lambda_2}}{\partial t}\right)\; ,\nonumber\\
K_{03}&=&-\lambda_0\lambda_3+\frac{1}{2} \left(\frac{\partial\lambda_0}{\partial\phi}+\frac{\lambda_3}{\partial t}\right)\; ,\nonumber\\
K_{12}&=&-\lambda_1\lambda_2+\frac{1}{2}\left(\frac{\partial \lambda_1}{\partial\theta}+\frac{\partial\lambda_2}{\partial r}\right)-\frac{  \lambda_2}{r}\; ,\nonumber\\
K_{13}&=&-\lambda_1 \lambda_3+\frac{1}{2} \left(\frac{\partial\lambda_1}{\partial \phi}+\frac{\partial\lambda_3}{\partial r}\right)-\frac{  \lambda_3}{r}\; ,\nonumber\\
K_{23}&=&-\lambda_2 \lambda_3+\frac{1}{2}\left(\frac{\partial\lambda_2}{\partial \phi}+\frac{\partial\lambda_3}{\partial \theta}\right)- \cot (\theta ) \lambda_3\; .
\eea

It is crucial to emphasize that none of these terms depend on $b(r)$. Then, the outcomes which follow from the non-diagonal terms will be valid for any choice of $b(r)$. On the other hand, in order for non-diagonal components of $K_{\mu\nu}$ to vanish, it seems reasonable to suggest:
\bea
\lambda_0=\lambda_2=\lambda_3=0\;  ,\label{lambdaform}
\eea
since otherwise it would be difficult to cancel the terms where they appear alongside metric  terms. This aligns with the methodology employed in \cite{Ayuso:2020vuu} \footnote{
However, it is possible to find other, more intricate, solutions if we take $\Phi'(r)=0$:
\begin{itemize}
    \item $\lambda_0\neq0$, $\lambda_1=0$, and  $\frac{\partial\lambda_0}{\partial r}=0$.
    \item $\lambda_0=\lambda_1=\frac{1}{-2 t - C_1(r - t)}$
\end{itemize}
}. Therefore, from now on, we will be considering the following form:
\bea
\lambda_\mu=(0,\lambda_1(r),0,0) \; , \label{lambdaconstraint}
\eea
Henceforth, all functions just depend on $r$ so we will not write explicitly this dependence in the equations, except in cases that might lead to confusion. The non-trivially satisfied field equations are these:
\bea
8\pi\rho=\frac{b'}{r^2}-\xi(\lambda^2_1-\lambda'_1)+\left(\frac{2\xi}{r}+\frac{\xi'}{2}\right)\lambda_1\; ,\label{rhoeq}
\eea
\bea
8\pi p_r=-\frac{b}{r^3}+\frac{2\xi}{r}\Phi'-\left(\frac{2\xi}{r}+\xi\Phi'\right)\lambda_1 \label{eqpr}\; ,
\eea
\bea
8\pi p_t&=&\frac{\xi'}{2r}+\left(\xi\left(\frac{1}{r}-\lambda_1\right)+\frac{\xi'}{2}\right)\left(\Phi'-\lambda_1\right)+\nonumber\\
&+&\xi\left(\Phi''+\Phi'^2-\lambda'_1\right)\; ,
\eea  
where
\bea
\xi=\left(1-\frac{b}{r}\right)\; .
\eea

Then, we have a system of three equations with six functions to be solved: $\rho(r)$, $p_r(r)$, $p_t(r)$, $\Phi(r)$, $b(r)$,  $\lambda_1(r)$, all of them dependent on $r$. In addition, note that the equation for $\rho(r)$ does not depend on $\Phi(r)$, which is associated with the tidal force exerted by the wormhole and the question of its potential traversabity. Moreover, we can recover the GR case by imposing $\lambda_1(r)=0$, leading to the customary equations for the space-time \eqref{MTst} in GR \cite{Morris:1988cz}.

\section{Analysis of the dissipative character of the theory}\label{V}

The equations of motion \eqref{eqmotion} allow to study the conservation of the matter energy-momentum tensor, thereby facilitating to examine the existence of dissipative effects. From a mathematical point of view, it can be viewed as:
\bea
\nabla_{\nu}K_{\mu}^{\;\;\nu}=8 \pi \nabla_{\nu}T_{\mu}^{\;\;\nu}\; . \label{eqconserv}
\eea
Hence, the presence of the $K_{\mu}^{\;\;\nu}$ tensor introduces the possibility of the non-conservation of the energy-momentum tensor. 

The $t$-component of the left-hand-side of Eq. \eqref{eqconserv} for the Morris-Thorne metric \eqref{MTst}, together with the constraint in the form of $\lambda_\mu$ as in \eqref{lambdaconstraint} resulting in the vanishing of the non-diagonal terms of $K_{\mu}^{\;\;\nu}$, reads:
\bea
\nabla_{\nu}K_{0}^{\;\;\nu}=\partial_\nu K_{0}^{\;\;\nu}+\Gamma^{\nu}_{ \nu\alpha}K_{0}^{\;\;\alpha}-\Gamma^{\alpha}_{\nu 0}K_{\alpha}^{\;\;\nu}=0\; ,
\eea
which automatically implies the conservation in time of the energy density.
\bea
\dot{\rho}=0 \; .
\eea
Indeed, this observation constitutes a significant result coming from the staticity of the metric.

However, the non-conservation effect does appear manifestly in the remaining components:
\bea
\nabla_{\nu}K_{r}^{\;\;\nu}=-\lambda_1 \left[\left(2 f \lambda_1+\Phi '^2+\Phi ''\right)\xi+f \xi ' \right]\label{conserK}\; ,
\eea
where
\bea
f=\frac{1}{r}+\frac{\Phi'}{2}\; .
\eea
From Eq. \eqref{eqconserv}, one must equate this result to the radial component of the covariant derivative of the energy-momentum tensor, which reads:
\bea
\nabla_{\nu}T_{r}^{\;\;\nu}=\partial_r p_r+\Phi' (p_r+\rho)+\frac{2}{r}\left(p_r-p_t\right)\label{conserT}\; .
\eea

Hence, the radial component of the (non)-conservation equation of the energy-momentum tensor \eqref{eqconserv} will vanish in the  $\lambda_1(r)=0$ case, which is evident from Eq. \eqref{conserK}. This makes sense as that condition reinstates GR, and  setting Eq. \eqref{conserT} equal to zero would yield the profile of $p_r(r)$ in a way similar to the Tolman–Oppenheimer–Volkoff (TOV) equation for compact objets. In the case with $\lambda_1(r)\neq 0$, the profile of $\partial_r p_r(r)$ will be altered by the contribution of $\nabla_\nu K_{r}^{\;\;\nu}$. However this modification is perfectly consistent with the calculation of the derivative of the equation of motion \eqref{eqpr}, as it should not come as a surprise. It means that there are two methods to calculate the derivative of $p_r(r)$. The first one consists in deriving Eq. \eqref{eqpr}, while the second one uses the radial (non)-conservation equation of the pressure, thus leading to an identical result. The meaning of this is that the solution of $p_r$ from Eq. \eqref{eqpr} already encapsulates the dissipative character of the theory, since its derivative does not correspond to the usual term we would obtain in GR.

The observational consequence of this is relevant. By measuring $\rho(r)$, $p_r(r)$ and $p_t(r)$, the dissipative character of the theory can be discerned given the interrelation between these three components. This important result extends beyond wormholes and applies to any object  described by the metric \eqref{MTst}, such as stars and other static bodies with spherical symmetry.

\section{Schwarzschild wormhole 
}\label{VI}

Searching for wormhole solutions is an exigent task that requires the adoption of some strategy. Specifying either $b(r)$ and/or $\Phi(r)$ is a typical one. In this section we will fix these functions to correspond to the Schwarzschild wormhole. The purpose is to elucidate some ideas and conditions of Lazo's theory exposed within a familiar and well established framework. Therefore, due to our interest in this extended theoretical framework, this case of a wormhole serves as a runway for new solutions. Furthermore, as we have already anticipated, the Schwarzschild wormhole is a particular case of the Morris-Thorne metric for the specific case where:
\bea
&&b(r)=b_0\; ,\label{mteq1}\\
&&\exp{2\Phi(r)}=1-\frac{b_0}{r}\; ,
\label{mteq2}
\eea
here $b_0$ being a positive constant. Recall that  Schwarzschild's metric was identified as a wormhole by Flamm \cite{Flamm, currentFlamm} just a year after Einstein's final formulation of his field equations. However, wormholes of this type  are essentially not traversable for three reasons \cite{Morris:1988cz}: 
i) the tidal gravitational force experienced within these wormholes is incredibly intense, to the extent that a traveler would not survive. This force is comparable in magnitude to the tidal force encountered at the horizon of a Schwarzschild black hole \cite{Misner:1973prb}; ii) 
Schwarzschild wormholes are dynamical, and their throats close so fast that there is not enough time for a traveler to pass through them \cite{wheeler1962geometrodynamics}; iii)
the presence of a past horizon, which is unstable to small perturbations, significantly complicates any attempts at achieving traversability \cite{10.1143/PTP.73.1401}.

\begin{figure}
    \centering
    \includegraphics[scale=0.6]{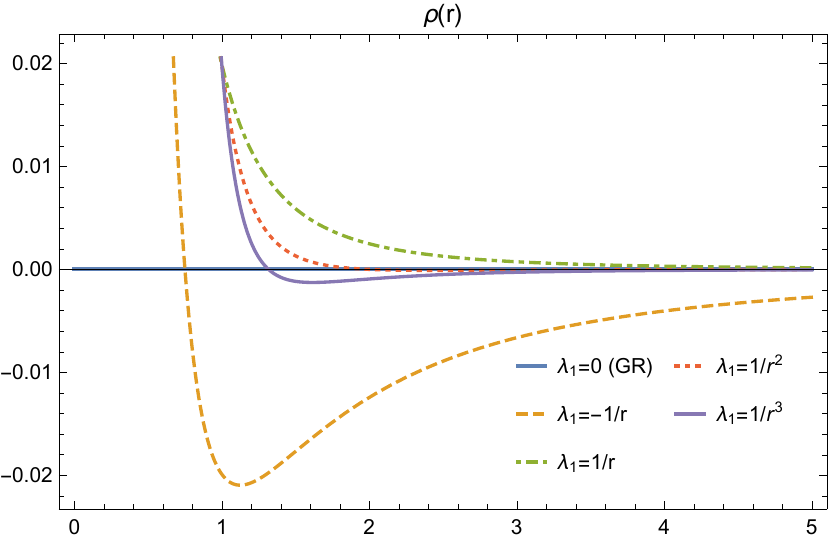 }
    \caption{Energy density profile for the Schwarzschild wormhole with various power-law choices of $\lambda_1$, where we have taken $b_0=1$. The case $\lambda_1=0$ is the case of GR since the new phenomenology dissapears recovering the Schwarzschild case as a vacuum solution with $\rho(r)=0$.}
    \label{SCHcase}
\end{figure}

Certainly, the Schwarzschild wormhole is very useful to address some relevant aspects. If we substitute Eqs \eqref{mteq1} and \eqref{mteq2} into our previous equation for $\rho(r)$, we obtain:
\bea
16 \pi  r \rho=\left(4-\frac{3 b_0}{r}\right)\lambda_1+&2 (b_0-r) 
\left(\lambda_1^2-\lambda_1'\right)\; . \label{lanrho}
\eea
Even though Lazo's theory offers nearly complete freedom regarding $\lambda_\mu$, we have already seen that the complete equations for a diagonal system imply the conditions of Eq. \eqref{lambdaconstraint}. In addition to this, the physical requirement $\lim_{r\to\infty}\rho(r)=0$ sets an important restriction. A straightforward option meeting these criteria is a power-law proposal such as $\lambda_1(r)\propto r^n$ with $n<0$. This choice satisfies the conditions discussed and  illustrated in Fig. \ref{SCHcase}. On the other hand, selecting $\lambda_1(r)=0$ effectively takes us  back to the GR scenario, wherein the Schwarzschild solution constitutes a vacuum solution. In this case, the energy density from Eq. \eqref{lanrho} simplifies to $\rho(r)=0$.

On the other hand, the equation for the radial pressure reads:
\begin{equation}
p_r=\frac{(3 b_0-4 r)}{16 \pi  r^2}\lambda_1\; .
\end{equation}

The latter is a useful result, as understanding energy conditions is crucial in the evaluation and validation of wormhole solutions. 
The Null Energy Condition 
reads:
\bea
\rho+p_r=\frac{(b_0-r) \left(\lambda_1^2-\lambda_1'\right)}{8 \pi  r}\geq 0\; .\label{NEC}
\eea
Note that the case $\lambda_1^2-\lambda_1'=0$ allows Eq. \eqref{NEC} to be satisfied. The solution of this differential equation is:
\bea
\lambda_1=-\frac{1}{r+c}\; ,
\eea

\begin{figure}
    \centering
    \includegraphics[scale=0.6]{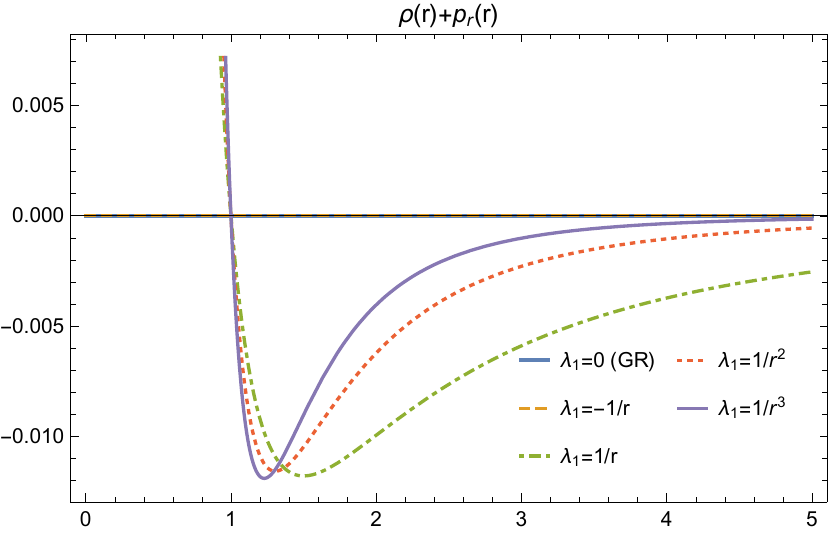 }
    \caption{$\rho(r)+p_r(r)$ profile for the Schwarzschild wormhole with various power-law choices of $\lambda_1$, where we have taken $b_0=1$. The case $\lambda_1=0$ is the case of GR since the new phenomenology dissapears.}
    \label{SCHcaseNEC}
\end{figure}

However, by examining Fig. \eqref{SCHcase} it becomes evident that $\rho(r)<0$ in that scenario. As a consequence, the Weak Energy Condition, which imposes a positive energy density in addition to Eq. \eqref{NEC}, will be breached (even though the NEC were satisfied). This violation is obvious as shown by:
\bea
\rho|_{\lambda_1=-\frac{1}{r}}=\frac{3 b_0-4 r}{16 \pi  r^3}\; .
\label{rho1Sch}
\eea

Other profiles for the study of the NEC with $p_r(r)$ are shown in Fig. \ref{SCHcaseNEC}. However, the case $\lambda_1(r)=-1/r$ is the only one among them that satisfies the NEC for $p_r(r)$. It is important to remind at this point that if the NEC condition is infringed, 
the rest of energy conditions will do as well. 

However, the study of the NEC cannot remain solely in the analysis with the radial pressure contribution; we must also study it under the use of tangential pressure. In this way, for the Schwarzschild wormhole, the NEC with the tangential pressure contribution is:
\bea
\rho+p_t=\frac{(2 r-3 b_0) \lambda_1}{16 \pi  r^2}\geq 0\; .\label{NEC2}
\eea
This result would lead us to consider a negative $\lambda_1(r)$ for $r<3b_0/2$ and a positive $\lambda_1(r)$ for $r>3 b_0/2$. An easy example is:
\bea
\lambda_1=\frac{r-3b_0/2}{r^2}\; .
\eea
Nevertheless, this kind of $\lambda_1(r)$ functions will violate the condition $\rho(r)+p_r(r)\geq0$ and consequently, the NEC is not fully satisfied for this case. It can be easily generalized that the NEC will always be violated beyond the particular cases presented here. First, from Eq. \eqref{NEC}, we saw that $\lambda_1'(r)$ must satisfy $\lambda_1'(r)\geq \lambda_1^2(r)$. Secondly, from Eq. \eqref{NEC2}, we saw that $\lambda_1(r)$ must change sign, which, along with the fact that it must vanish at infinity, leads us to the conclusion that there will always be some point where its derivative vanishes. Consequently, the only function that satisfies all of this is $\lambda_1(r)=0$, reverting to the well-known case of GR.

For those reasons, this wormhole puts us on the path to studying new configurations of the Morris-Thorne metric, devoid of problems such as lack of traversability or negative energy density.

\section{Morris-Thorne wormhole solutions from positive energy density}\label{VII}

Typically, when working in the field of gravitation, two {\it a priori} different paths are  followed. The first way, a spacetime of potential interest is chosen an the type of matter and energy that can sustain it are explored using gravitational equations. In fact, this is what was done in the previous section. The second way, one considers specific matter distribution and types, and starting from there, the spacetime is reconstructed again using gravitational equations.

In this section, we will embark on a combination of both. First, we craft
an energy profile with a series of desirable properties. This will allow us to reconstruct the shape function $b(r)$ for a metric of the form in Eq. \eqref{MTst}.  Additionally, we will demand the solutions found depict traversable wormholes, so we will set $\Phi'(r)=0$. Note that  this does not affect the obtained $\rho(r)$ profile, but it does modify the pressures. This kind of spacetimes are usually called ultrastatic \cite{Sonego:2010vy}.

Besides, we start focusing on the particular and easier case of $\lambda_1(r)=-1/r$ that lets us get a simpler equation of $\rho(r)$, as a first approach, and motivated by the Schwarzschild case where this form of the function managed to satisfy a part of the NEC.
\bea
\rho=\frac{3 (r b)'-4 r}{16 \pi  r^3}\; .\label{fappr}
\eea

Accordingly, our aim will be to construct wormhole solutions, ensuring that $\rho(r)$ is positive at the very least.  Keeping the latter purpose in mind, we will adopt the following expression for the energy density:
\bea
\rho =\frac{\beta+4s(r)-4r}{16 \pi r^3}\; ,\label{rhoproposal}
\label{rho1}
\eea
where $\beta$ is a positive constant and $s(r)$ is a function of $r$. Then, $\rho(r)$  can easily be forced to be positive throughout $r$ with an appropriate choice of the function $s(r)$.  This choice for $\rho(r)$ is inspired by the Schwarzschild case solution (Eq. \eqref{rho1Sch}), but with a generalized term for $b_0$. By equating equations \eqref{fappr} and \eqref{rhoproposal}, we obtain a new differential equation whose solution will give us the shape function of the metric:
\bea
3 (r b)'-\beta-4 s=0\; ,\label{difeq}
\eea
Clearly, for a generic $s(r)$ the solution will take the form:
\bea
b=\frac{\beta+2r}{3}+\frac{4}{3r}\int (s(r)-r)dr\; .
\eea

By establishing the form of $s(r)$, and consequently that of $\rho(r)$, we open up the possibility of solving the differential equation \eqref{difeq}. However, our preference is for $\rho(r)$ to generally be an asymptotically vanishing positive continuous function.

This compact way to formulate our generic solutions allows to rewrite neatly some of the relevant conditions. The flaring out condition translates into 
\bea
b'(r_0)=-1+\frac{\beta+4 s(r_0) }{3 r_0}<1
\eea

Following this methodology, one is able to find a multitude of solutions. In order to analyze this, let us show an example under the following specific choice: 
\bea
s=r\left(1+\frac{\alpha}{\left(1+r^2\right)^{n/2}}\right)\; , \label{proposal1}
\eea
where $\alpha$ and $n$ are both positive constants.
This expression for $s(r)$ comprises two terms. The linear term is selected in such a way that when the non-linear term is disregarded, $\rho(r)$ is precisely proportional to $\beta$. If this parameter is positive, all three conditions outlined above regarding $\rho(r)$ are met. 

\begin{figure*}
    \includegraphics[scale=0.6]{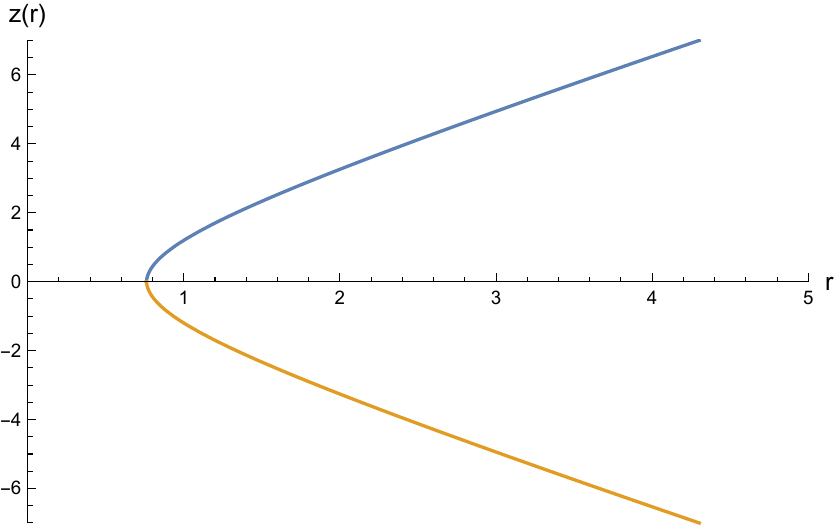}
    \includegraphics[scale=0.6]{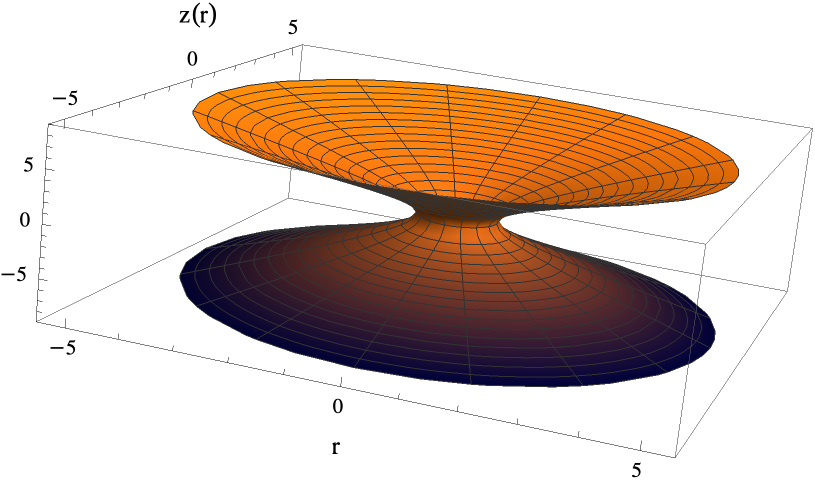}
    \caption{Embedding of the wormhole given by Eq. \eqref{br1} for $\alpha=0.1$, $\beta=0.1$ and $n=1$ in 2D (left panel) and 3D (right panel). This is characterized by the conicity of the spacetime, that can be better observed in the 2D graphic.}
    \label{Figembedding1}
\end{figure*}

However, the contribution of the non-linear term is secondary; it introduces a bell-shaped adjustment 
which, for $n>1$, ensures the attenuation of $\rho(r)$ as $r$ increases. This expression for $s(r)$ enables us to derive a positive $\rho(r)$, as the term $-4r$ in Eq. \eqref{rho1} is cancelled by the linear term in Eq. \eqref{proposal1}. Following this selection our solution is characterized by:
\begin{equation}
\label{br1}
b=\frac{\beta+2r}{3}+ 
\begin{cases}
\displaystyle\frac{4 \alpha}{3 (2-n)
   r}\left(1+r^2\right)^{1-{n}/{2}}\; ,\,  &\text{if } n \neq 2\; , \\
   \\
  \displaystyle \frac{ 2\alpha }{ 3r}\log \left(1+r^2\right)\;,\,  &\text{if } n=2\; ,
\end{cases}
\end{equation}
where an arbitrary integration  constant has been set to zero. Subsequently, we can numerically solve the expression for the embedding of the Morris-Thorne metric, as previously mentioned in Eq. \eqref{embeddingequation0}, for this particular form of $b(r)$. The results in both 2D and 3D are presented in Fig. \ref{Figembedding1}.

Additionally, to ensure the presence of a throat in the wormhole, we need to study the flaring out condition expressed in Eq. \eqref{floutcondition} for the general case, and which is specified as follows for the solution found:
\bea
    \frac{1}{6\; b^2} \left(\beta-\frac{4 \alpha  \left(n r^2+2\right)}{(n-2) r\left(1+r^2\right)^{n/2}}\right)>0\; ,  &&\text{if } n \neq 2\nonumber\\
    \\
    \frac{1}{6\; b^2} \left(\beta-\frac{4 \alpha r}{1+r^2}+\frac{4 \alpha \log \left(1+r^2\right)}{r}\right)>0, && \text{if } n=2\nonumber
\eea
Then, the inequality is satisfied for $\alpha>0$, $\beta>0$, and $n\leq2$  $\forall r > 0$. Nevertheless, examining the left-hand side of the inequality, we can see that it approaches zero as $r$ tends to infinity. 
This behaviour stems from the conical nature of this type of wormhole, as previously mentioned, and we will now delve into it more extensively.  

The introduction of the equal sign in the inequality \eqref{floutcondition} was originally proposed by Hochberg and Visser \cite{Hochberg:1998qw}, leading to its designation as the simple flare-out condition. Conversely, when the inequality is presented without the equal sign, as in Eq. \eqref{floutcondition}, it is termed the strong flare-out condition \cite{Kim:2013tsa}. The occurrence of the second derivative becoming zero can be interpreted as the first derivative being a constant at infinity, implying $z\propto  r$ in the embedding diagram. This observation highlights the conical nature of the wormhole, therefore suggesting the absence of asymptotic flatness.

The asymptotic flatness is the characteristic that a spacetime ends up being Minkowski at spatial infinity. For the space-time we study here, it can be translated into the following condition:
\bea
\lim_{r\rightarrow \infty}\left(1-\frac{b}{r}\right)=1\; . \label{limit}
\eea

However, it is not satisfied in general for the wormhole solutions we have found in Eq. \eqref{br1} as here is a breakdown of the possibilities:
\begin{itemize}
    \item If $n<0$, the limit diverges, deviating from the desired flatness condition. This is in agreement with the fact that the energy density also diverges at infinity as can we check from Eq. \eqref{proposal1}. 
    \item When $n=0$, the limit evaluates to $(1-2\alpha)/3$. In this case, satisfying the flatness condition is possible if $\alpha=-1$, but it comes at the expense of losing positivity in the energy density.
    \item For $n>0$, the limit converges to $1/3$, indicating again an inconsistency with flatness. This outcome leaves out the expectation of a flat spacetime at spatial infinity, characteristic of a conical spacetime.
\end{itemize}

These findings indicate that the solution of the wormhole with this particular choice of $\rho(r)$ and $\lambda_1(r)$ exhibits a  deficit angle, implying that the spacetime is asymptotically conical \cite{Jusufi:2018waj}.

As this will be a recurring outcome in our investigation, we aim to offer a more pedagogical insight that complements what has already been discussed regarding the flaring out condition, aiding in the comprehension of the conicity of the spacetime under consideration. To achieve this, we will focus on the specific scenario where $n>0$ but $n\neq 2$, and we will examine a constant-time slice projected onto the $\theta=0$ plane. Consequently, for sufficiently large $r$, we can approximate the metric as:
\bea
ds^2=3 dr^2+r^2d\phi^2\; .
\eea

Introducing a redefined radial coordinate $l=\sqrt{3}r$ and an angular coordinate $\tilde\phi=\phi/\sqrt{3}$, we obtain
\bea
ds^2= dl^2+l^2d\tilde\phi^2\; .
\eea
This line element resembles that of 2D Minkowski spacetime expressed in polar coordinates, with $l$ serving as the radius.  We wonder then where does the conicity of spacetime manifest itself. The answer is clear: the domain of the angular coordinate $\tilde\phi$ is not $[0,2\pi]$, but rather $[0,2\pi/\sqrt{3}]$, and therefore the deficit angle that the metric will have is the following:
\bea
\psi=2\pi\left(1-\frac{1}{\sqrt{3}}\right)\; .
\eea

A well-known consequence of this spacetime type is that the length of a circumference with radius $l$ is no longer $2\pi l$, but rather
\bea
\text{length}=\int_{0}^{\frac{2\pi}{\sqrt{3}}}{l d\tilde\phi}=\frac{2\pi}{\sqrt{3}}l
\eea

It is possible to revert this situation with a slight modification:
\bea
s=r\left(\gamma+\frac{\alpha}{\left(1+r^2\right)^{n/2}}\right)\;.
\eea
where $\gamma$ is a constant. This case transforms the expression  of $b(r)$ into:
\bea
b=\frac{\beta r+2 \alpha \log \left(1+r^2\right)+2 \gamma  \left(1+r^2\right)}{3 r}\; ,
\eea
and the lhs of Eq. \eqref{limit} reads now $1-2\gamma/3$. Consequently, for $\gamma=0$ the flatness condition is recovered, but it can be seen that the  toll to be paid for this is that the   energy density becomes negative is some region of the space again.

It is possible to generalize this conclusion and check that this conicity  condition will be a consequence of the choice of $\lambda_1(r)=-1/r$ for a positive $\rho(r)$. To demonstrate this, let us assume that the flatness condition is satisfied and take the spatial infinity limit of Eq. \eqref{difeq} divided by 
$r$:
\bea
\lim_{r\rightarrow\infty}\left(3 b'-\frac{4 s(r)}{r}\right)=0 \label{cond}
\eea

It seems logical to think that if we are looking for a shape function that grows (or decreases) more slowly than $r$, so that the ratio of both vanishes at infinity, we can assume that $b'(r)$ will become zero on this same limit\footnote{Of course, we can find the exception of fluctuation functions as trigonometric examples, but then the limit will fluctuate as well and we can find points where $b'(r)$ vanishes.}. Then, we can suppress the first term and draw the conclusion that $s(r)$ cannot have linear terms in $r$ as \eqref{cond} would not be satisfied. However, it would imply the reappearance of the negative term in $\rho$ (Eq.  \eqref{rhoproposal}) vanishing the possibility to construct a positive energy density throughout the space.

Thus, in this specific context of wormhole construction, it can be inferred that a positive energy density is associated with the emergence of conical singularities.

In addition, we can examine the $\rho(r)+p_r(r)$ part of the NEC. For $\Phi'(r)=0$, it simplifies considerably and the expression associated with the NEC reads
\begin{equation}
\rho+p_r=
\begin{cases}
  - \frac{1}{16 \pi}\left(\frac{4 \alpha  \left(2+n r^2\right)}{(2-n)\left(1+r^2\right)^{n/2}r^4}+ \frac{\beta}{r^3} \right)\; , & \text{if } n \neq 2 \\
   \\
    -\frac{1}{16 \pi}\left(\frac{4 \alpha}{r^4} \left(\log \left(r^2+1\right)-\frac{r^2}{r^2+1}\right)+\frac{\beta}{r^3}\right)\;, & \text{if } n=2.
\end{cases}
\end{equation}
Numerical examination shows this is typically violated regardless of the value of the (positive) parameter $n$.

Due to this, one might reasonably wonder if the violation of the NEC is a consequence of the theory when $\lambda_1(r)=-1/r$ or if it only occurs in this particular solution. To do so, we will study $\rho(r)$ and $p_r(r)$ at the throat of the wormhole. Taking into account that the flaring out condition can be translated into $b'(r_0)<1$, the energy density will be bounded from above at the throat as 
\bea
\rho(r_0)=\frac{3 b'\left(r_0\right)-1}{16 \pi  r_0^2}<\frac{1}{8 \pi  r_0^2}\; .
\eea
This matches the limit derived in GR \cite{bookLobo} leaving open the possibility of a positive value. Remarkably, this equivalence holds true even with the structure being significantly more intricate, enabling greater solution complexity. Similar circumstances arise when exploring the radial pressure for $\lambda_1(r)=-1/r$:
\bea
p_r=\frac{3 r (r-b) \Phi '-3 b+2 r}{8 \pi  r^3}\; .
\eea
Thereafter, when evaluated at the throat, yields
\bea
p_r(r_0)=-\frac{1}{8 \pi  r_0^2}<0\; ,
\eea
coinciding once again with the value obtained in GR\footnote{Yet another match can be identified when studying the difference between the radial pressure from Lazo's theory and the one from the GR:
\bea
p^{Lazo}_r-p_r^{GR}=\frac{(r-b) \left(r \Phi '+2\right)}{8 \pi  r^3}.
\eea
It is straightforward to see that both values just coincide at the throat, i.e. when $b(r)=r$.}. Furthermore, the NEC must be violated at the throat, as
\bea
\rho(r_0)+p_r(r_0) < 0\ ,
\eea
as long as $\lambda_1(r)=-1/r$ is chosen, and regardless of the specific wormhole solution we may find. Following the procedure of this section, it is easy to discover other solutions for different choices of the function $s(r)$ with the same characterization.

\section{General study of the NEC}\label{VIII}

In the previous section, we have described a methodology to build wormhole-type solutions with a positive energy density when the free parameter $\lambda_1(r)$ is fixed. However, when we set $\lambda_1(r)=-1/r$ to simplify the calculations, the obtained solutions will have two problems: 
the first is the violation of the NEC, and the second is the lack of asymptotic flatness or conicity of the spacetime.

In this section, we will try to shed some light on the NEC in wormholes constructed from this theory, relaxing for this purpose the choice of $\lambda_1(r)$. As in the previous section, we start this analysis particularizing it  at the throat and considering $\Phi'(r)=0$: 
\bea
\rho(r_0)+p_r(r_0)=\frac{1-b'(r_0)}{8\pi r_0}\left(\frac{\lambda_1(r_0)}{2}-\frac{1}{r_0}\right)\; .\label{NECth}
\eea

Considering the flaring out condition,  $b'(r_0)<1$, the sign of the rhs of Eq. (\ref{NECth}) will depend just on the term in parentheses. Consequently, if it is positive, this part of the NEC will be satisfied at the throat, otherwise, if this term is negative, NEC will not be satisfied any longer. This conclusion is consistent with the previous result for $\lambda_1(r)=-1/r$, which we had already shown does not satisfy the NEC. However, the case $\lambda_1(r)=\alpha/r$ with $\alpha\geq 2$ will satisfy this part of the NEC, at least in the throat, but we cannot guarantee that it will be satisfied throughout the entire space.

The radial NEC in its most general form, merely combining metric Eq. \eqref{MTst} and the $\lambda_mu$ structure in Eq. \eqref{lambdaconstraint} reads:
\bea
\rho+p_r&=&-\frac{1}{8 \pi}\left[\xi'\left(\frac{1}{r}-\frac{\lambda_1}{2}\right)\right.\nonumber\\
&+&\left.\xi\left(\lambda_1^2-\lambda_1'-2\left(\frac{1}{r}-\frac{\lambda_1}{2}\right)\Phi'\right)\right]\; .\label{NECgen}
\eea
However, we should consider some characteristics that the functions $\xi(r),\; \Phi(r)$ must satisfy to represent a wormhole. Firstly, the function $\xi(r)$ must be such that it is monotonically increasing and intersects the r-axis at only one point, where the throat will be located. Secondly, $\Phi(r)$ must go to zero and $\xi(r)$ must approach $1$ at spatial infinity to recover asymptotic flatness. Additionally, it seems logical to assume that $\lambda_1(r)$ must be a function that vanishes at infinity. As it could not be otherwise, if we bring it to the throat ($\xi(r_0)=0$ and $\xi'(r_0)>0$), we recover Eq. \eqref{NECth}, along with the condition $\lambda_1(r_0)>2/r_0$, that could be extended in the whole space as $\xi'(r)>0$:
\bea
\frac{1}{r}-\frac{\lambda_1}{2}<0\; .
\eea
With this, we  ensure that the term accompanying $\xi'(r)$ is not only positive at the throat, but also for all $r$. 

We then move on to analyze the term accompanying $\xi(r)$. In this second term, we have $\xi(r)$, which will be positive. The term $\lambda_1^2(r)-\lambda_1'(r)$ will also be positive, as one is a square and the other is the derivative of a function that is larger than $2/r$ at the throat
and ends at zero at infinity\footnote{Of course, this is only true if we consider a monotonically decreasing function, but even if it is not, it must be decreasing at certain points in space, so the analysis remains valid.}. Moreover, the term $(1/r-\lambda_1/2)$ will be negative, as we have previously observed. This leads us to require that $\Phi'(r)$ be negative and such that:
\bea
\lambda_1^2-\lambda_1'-2\Phi'\left(\frac{1}{r}-\frac{\lambda_1}{2}\right)<0 \label{ineq}\; .
\eea

However, this will imply a $\Phi(r)$ that does not tend to zero at spatial infinity. To demonstrate this, we take the limiting case where $\lambda_1(r)=\alpha/r$ with $\alpha>2$ and substitute into Eq. \eqref{ineq}, obtaining:
\bea
\Phi'<-\frac{(\alpha^2+\alpha)}{\alpha-2}\frac{1}{r}=-\frac{c}{r}\; ,
\eea
i.e. this inequality implies that $\Phi'(r)$ must decrease slowly enough for $\Phi(r)$ not to approach zero at infinity, thereby breaking the wormhole structure.

However, this development has an exception. This occurs when, at the throat, the second term of Eq. \eqref{NECgen} does not vanish because (even though $\xi(r)$ remains zero), the term $\Phi'(r)$ tends to diverge (as $1/0$) at the throat and both terms compensate each other. This is exactly what happens in the Schwarzschild case, where we found a possibility to satisfy the radial part of the NEC for $\lambda_1(r)=-1/r$ at the cost of the wormhole being non-traversable due to its dynamics and tidal forces. This can be extended to more complex wormhole profiles, as for this second term of Eq. \eqref{NECgen} not to vanish at the throat, it must always be due to a divergent $\Phi'(r)$. Therefore, the conclusion of non-traversability will be the same as in the Schwarzschild case.

\section{Conical character}\label{IX}

Finally, in this section, we will examine whether the wormholes constructed from these theories are bound to be conical whenever we impose a positive energy density, or if, on the contrary, we can build solutions that satisfy the flatness condition even when $\rho(r)>0$.

For this purpose, we take the general equation for $\rho(r)$ and exactly as we did in the previous section we do not  fix $\lambda_1(r)$. 
To ensure the flatness condition from the beginning, we will adjust the form of the shape function. Specifically, we will consider the simplest case where $b(r) = b_0$. In this scenario, the  $g_{rr}$ component of the metric will be identical to that of the Schwarzschild metric.
Moreover, let us recall that the form of $\Phi(r)$ will not affect the profile of $\rho(r)$ in any way. With these hypotheses we can rewrite Eq. \eqref{rhoeq} as:
\bea
\rho=\frac{1}{8\pi r^2}-\frac{\xi}{8\pi}\left(\left(\frac{1}{r}-\lambda_1\right)^2-\lambda_1'\right)-\frac{\xi'}{8\pi}\left(\frac{1}{r}-\frac{\lambda_1}{2}\right)\; ,\nonumber\\ \label{eqrhoN}
\eea

Similarly to the discussion in Section \ref{VII}, where the profile of $\rho(r)$ is enforced to be positive, we will manually impose a positive energy density. Thus, we will propose \bea
\rho=\frac{1}{8\pi r^n}\label{rn}\; ,
\eea as this is a simple choice that vanishes at infinity for $n>0$.

The use of this profile for $n \leq 2$ leads to an expression for $\lambda_1(r)$ which diverges at spatial infinity. However, cases with $n > 2$ are numerically solvable and lead to a  $\lambda_1(r)$ shape which vanishes at infinity. For the numerical solution, an initial condition must be provided. This will be given by the value that $\lambda_1(r)$ must take at the throat, which can be deduced by equating \eqref{eqrhoN} and \eqref{rn} at the throat:
\bea
\lambda_1(r=b_0)=\frac{2}{b_0^{n-1}}\; .
\eea

\begin{figure}
    \includegraphics[scale=0.6]{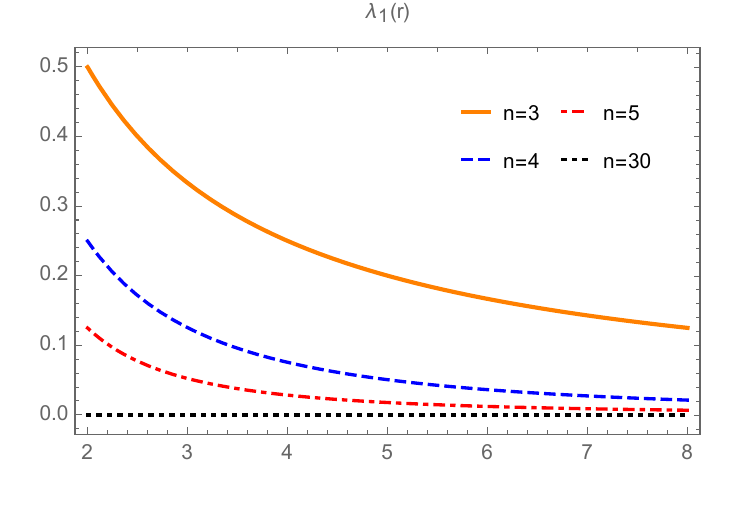}
    \caption{
    Numerical profiles of $\lambda_1$ obtained from equating Eq. \eqref{eqrhoN} with Eq. \eqref{rn} for different values of $n$, with $b_0=2$. All of them are associated with a positive energy density.}
    \label{figurerho}
\end{figure}

In Fig. \ref{figurerho}, we show the different profiles for $\lambda_1(r)$ depending on the value of $n$. Consequently, the theory allows for the construction of wormholes with a positive $\rho(r)$ profile that satisfy the flatness condition. However, in general, these solutions will involve convolved expressions for $\lambda_1(r)$ and therefore we have just explored the issue numerically.

This conclusion is consistent with the results obtained in Section \ref{VI}, where we had already found cases with a positive $\rho(r)$ for an analytical $\lambda_1(r)$. Nevertheless, while in that section the initial approach was to provide a $\lambda_1(r)$ from which we derived the $\rho(r)$ profile, in this section, the philosophy of the problem has been the opposite. That is, we enforce a given $\rho(r)$ and from it construct a supporting $\lambda_1(r)$. In fact, the case $n=3$ in this section becomes analytical and corresponds to the result $\lambda_1(r)=1/r$ from Section \ref{VI}.

\section{Summary and conclusions}\label{X}

We have studied traversable wormhole solutions in the framework of action-dependent Lagrangian theories, using for that purpose the Morris-Thorne metric. More specifically, we have studied the modified gravity theory with the Lagrangian density: $\mathcal{L}=R-\lambda_\mu s^\mu$,  where $s^\mu$ is the action-density field which names the theory. Then, from the symmetries of the space-time and the diagonal energy-momentum tensor, some constraints on $\lambda_\mu$ emerge which lead us to consider the case where this four vector acquires the form $\lambda_\mu=(0,\lambda_1(r),0,0)$. 

Under this constraint, we have studied the modification of the equation that relates the energy density with the pressure (in a similar way to the TOV equation) in Eq. \eqref{conserK}. This modification emerges from the dissipative character of the theory, or equivalently in this case, due to the covariant derivative of the energy-momentum tensor not being zero. Then, measuring the relation between these components, one is able to find new phenomenology associated with the theory.

Subsequently, we have studied the well-characterized Schwarzschild wormhole as a starting point and as a particular case of the Morris-Thorne metric. In doing so, we have found that under the particular choice of $\lambda_1(r)=-1/r$, it is possible to satisfy the radial NEC, but not the traversable NEC, although this results in a negative energy density. Furthermore, this choice of $\lambda_1$ greatly simplifies the calculations, leading us to use it for constructing new wormhole solutions with positive energy density.

However, these new wormholes constructed with $\lambda_1(r)=-1/r$ are bound to have two fundamental problems. The first is that they will not satisfy the NEC, and the second is that they do not satisfy the flatness condition, resulting in conical wormholes with an associated deficit angle. The violation of NEC in this case, where a positive $\rho(r)$ has been imposed, means that in the  inertial frame of reference comoving with the fluid, the energy density that supports the wormhole is positive, although an observer who traverses the throat with a velocity close to the speed of light could measure a negative energy density. 

Finally, we have shown that, in general, under a generic $\lambda_1(r)$, the NEC will always be violated, except in extreme cases that do not allow for the construction of a traversable and stable wormhole, as we have mentioned in the Schwarzschild case. However, the issue of conicity can be resolved for more general $\lambda_1(r)$'s that restore asymptotic flatness, while being supported by a positive $\rho(r)$.

\section*{Acknowledgments}

IA and RL are supported by the Basque Government Grant IT1628-22, by Grant PID2021-123226NB-I00 (funded by MCIN/AEI/10.13039/501100011033 and by “ERDF A way of making Europe”).
This article is based upon work from COST Action CA21136 Addressing observational tensions in cosmology with systematics and fundamental physics (CosmoVerse) supported by COST (European Cooperation in Science and Technology). 


\bibliography{Bib}



\end{document}